# AEcroscoPy: A software-hardware framework empowering microscopy toward automated and autonomous experimentation

Yongtao Liu,[1*] Kevin Roccapriore,[1] Marti Checa,[1] Sai Mani Valleti,[2] Jan-Chi Yang,[3] Stephen Jesse,[1] Rama K. Vasudevan[1†]

[1] Center for Nanophase Materials Sciences, Oak Ridge National Laboratory, Oak Ridge, Tennessee, United States of America
[2] Bredesen Center for Interdisciplinary Research, University of Tennessee, Knoxville, Tennessee, United States of America
[3] Department of Physics, National Chiao Tung University, Taiwan

---

[*] liuy3@ornl.gov
[†] vasudevanrk@ornl.gov

**Abstract**

Microscopy, in particular scanning probe and electron microscopy, has been pivotal in improving our understanding of structure-function relationships at the nanoscale and is by now ubiquitous in most research characterization labs and facilities. However, traditional microscopy operations are still limited largely by a human-centric click-and-go paradigm utilizing vendor-provided software, which necessarily limits the scope, utility, efficiency, effectiveness, and at times reproducibility of microscopy experiments. Here, we develop a coupled hardware-software platform that consists of a field-programmable gate array (FPGA) device, with LabView-built customized acquisition scripts, along with a software package termed AEcroscoPy (short for Automated Experiments in Microscopy driven by Python) that overcome these limitations and provide the necessary abstractions towards full automation of microscopy platforms. The platform works across multiple vendor devices on scanning probe microscopes and scanning transmission electron microscopes. It enables customized scan trajectories, processing functions that can be triggered locally or remotely on processing servers, user-defined excitation waveforms, standardization of data models, and completely seamless operation through simple Python commands to enable a plethora of microscopy experiments to be performed in a reproducible, automated manner. This platform can be readily coupled with existing machine learning libraries as well as simulations, to provide automated decision-making and active theory-experiment optimization loops to turn microscopes from characterization tools to instruments capable of autonomous model refinement and physics discovery.

**Introduction**

Automated and autonomous research is an emerging topic in many scientific fields, such as materials science,[1, 2] life science,[3] drug discovery,[4] climate science, and astronomy. Automated research leverages technology and machinery to perform experiments automatically to improve efficiency and precision, enabling round-the-clock operations. Autonomous research takes automated labs a step further, by using machine learning to assist in analysis and decision-making, enabling autonomous operations that can lead to accelerated discoveries.[5, 6]

Worldwide, there have been several notable examples of automated and autonomous systems at both research and industrial laboratories, such as those of the Emerald Cloud lab, as well as the efforts to 'digitize' chemistry from the Cronin group, and numerous others around the world.[7-12] Similarly, efforts to identify novel material growth methods to optimize properties have been implemented on 3D printing setups[13] and nano-synthesis methods[14-16]. Although most synthesis and characterization platforms are still heavily dependent on human operations, automation of sub-tasks is by now routine across many platforms. To date, however, automation of advanced characterization tools such as scanning probe and electron microscopes has remained limited, and the prevailing paradigm is that of an expert user 'driving' the tool and handling virtually all operational decisions.

Both scanning probe microscopy (SPM) and scanning transmission electron microscopy (STEM) have proved pivotal to expanding our knowledge of materials at the nanoscale, enabling researchers to determine structure-property relationships by combining high-resolution structural imaging with a wide variety of spectroscopic measurements to provide insights into the chemical features, electronic structure, and functional properties of the sample.[12, 17] The key advantage of microscopy is the ability to correlate the microstructural features from imaging mode with the functional properties from the spectroscopic mode that can also be measured in a spatially resolved manner.

Although SPM and STEM are ubiquitous across academic and industrial laboratories, the operation of these tools remains reliant on experienced researchers for manual operation, data analysis, and decision-making. For instance, in Piezoresponse Force Microscopy (PFM) and spectroscopy measurements of ferroelectric thin films,[18] the PFM image measurement region must be determined by human operators via overview scans; subsequent to a PFM image scan, a decision of the spectroscopy measurement locations is made by researchers based on observed domain structures in PFM image, along with the researchers' interest and prior knowledge. The specific commands that need to be sent to the instrument to enable the execution of this workflow are vendor-specific; this introduces significant cost as operators need to be re-trained whenever working on a different platform. Moreover, each vendor-provided software has limited (if any) Application Programming Interface (API) to enable automated control of microscope operations, further complicating the process.

As such, until now, executing workflows in microscopy experiments has mostly relied on the manual operation of vendor-given software by experienced researchers, which can greatly limit the efficiency, scope, effectiveness, and reproducibility of microscopy experiments. These limitations can be overcome by automated and autonomous experiments (AE) that execute experiment workflows in an automated manner, where workflows contain operation elements in sequence that define the experiment. Automation of this workflow first requires the development of software and hardware abstractions that can be readily applied to multiple instruments and platforms, i.e., cross-platform API, whilst allowing the user to use a common experiment hyper-language that describes the workflow with the necessary detail to implement on any hardware device. The hyper-language enables the user to focus on the microscope workflow, and iterate on experimental design, as opposed to spending time on local optimization of the implementation on the specific microscope employed. The cross-platform API provides the microscope community opportunities to share experiment protocols and implement methods in new microscope instruments with lower barriers. One example of such developments is the pycroManager software for universalizing control of optical microscopy setups,[19] which consists of a high-level programming interface exposed to the user, combined with low level hardware-specific blocks to optimally run experiments based on the user input provided. Another example of hyper-language development is the PyLabRobot, a cross-platform interface that offers a universal set of commands capable of programming diverse liquid-handling robots.[20] However, to date this concept has not been applied to the generic SPM and STEM, despite both sets of instruments essentially operating in similar manner, i.e, the user manipulates the trajectory of a scanning probe (physical in SPM, or focused beam of electrons in STEM) and acquires images, and performs spectroscopy measurements at different locations or under different conditions.

    Here, we develop a coupled hardware-software platform that consists of a field-programmable gate array (FPGA) device, with LabView-built customized acquisition scripts, along with a software package termed AEcroscoPy (short for Automated Experiments in Microscopy driven by Python) that provide the necessary abstractions towards automation of microscopy platforms. The platform works across multiple vendor devices on SPMs and STEMs. It enables customized scan trajectories, processing functions that can be triggered locally or remotely on the cloud, user-defined excitation waveforms, standardization of data models, and completely seamless operation through simple Python commands to enable a plethora of microscopy experiments to be performed in a reproducible, automated manner. This platform can be readily coupled with existing machine learning libraries as well as simulations, to provide automated decision-making and online theory-experiment optimization loops to turn microscopes from characterization tools to instruments capable of autonomous model refinement and physics discovery.

# 1. Hardware-software platform

Figure 1 shows the hardware-software platform that describes the AEcroscoPy system. A hardware device (FPGA) is used to output signals that are fed into the controller of the microscope platform and takes as input one or multiple channels of data through the same FPGA device. The first two output signals are used to control the position of the probe, enabling customized scan trajectories. The subsequent two output signals can be used to e.g., apply customized waveforms through the SPM tip for voltage-based spectroscopies in SPM, or e.g., used to modulate other excitations such as optical (laser), thermal (heated stages), etc. The FPGA device used is the National Instruments USB-7856 multifunction reconfigurable I/O. Note that in the case of band-excitation measurements, we combine the FPGA device with another National Instruments data acquisition card (PXIe-6124) to enable customized band-excitation spectroscopic measurements to be performed. In this case the FPGA works in concert with the additional DAQ card via hardware trigger signals for synchronization, for example to realize band-excitation waveforms to be sent to the tip during customized scan trajectories. To date we have been able to use this to support four SPM platforms (Asylum/Oxford Instruments, Bruker, Nanosurf and Nanonis (for STM)) as well as one STEM platform (Nion Swift).

The software portion of the AEcroscoPy system, as shown in Figure 1, consists of Labview-written executable virtual instrument (VI) to control the FPGA and enable rapid on-the-fly data visualization, as well as a fully featured python package (called the AEcroscoPy package) to allow users to write code in python to execute experiments without specific reference to the underlying hardware SPM/STEM system. The AEcroscoPy package utilizes the python for Windows (pywin32) package to access the variables and data stores within the Labview executable. It should be noted that the design of the Labview VI is such that all variables are exposed, enabling full automation through python, i.e., every function that can be accessed through the Labview executable can similarly be accessed through AEcroscoPy. There are two main Labview VIs in AEcroscopy: one for SPM control and voltage-spectroscopy, termed 'BEPyAE' and the other for customized scanning routines ('PyScanner'), screenshots of BEPyAE and PyScanner are shown in Figure S1. Both are often needed in SPM, but only the latter is required for STEM. Screenshots of the two VIs are provided in the supplementary. Different backends require either different triggering setups, or specialized functions that enable customized control of that instrument. These are too numerous to list here, but each case is unique, and the FPGA can be triggered via one or two-way triggers from the instrument controller.

From the user's perspective, the coding is done through the functions available in AEcroscoPy, or (if the required function is not available) by writing appropriate python code to alter parameters within the Labview executable. This code is not dependent on the specific instrument, but it should be noted that in certain instances only some types of experiments can be performed depending on the instrument, due to either hardware limitations (e.g., scan size ranges or availability of temperature stages), a lack of instrument control APIs by the vendor, or both.

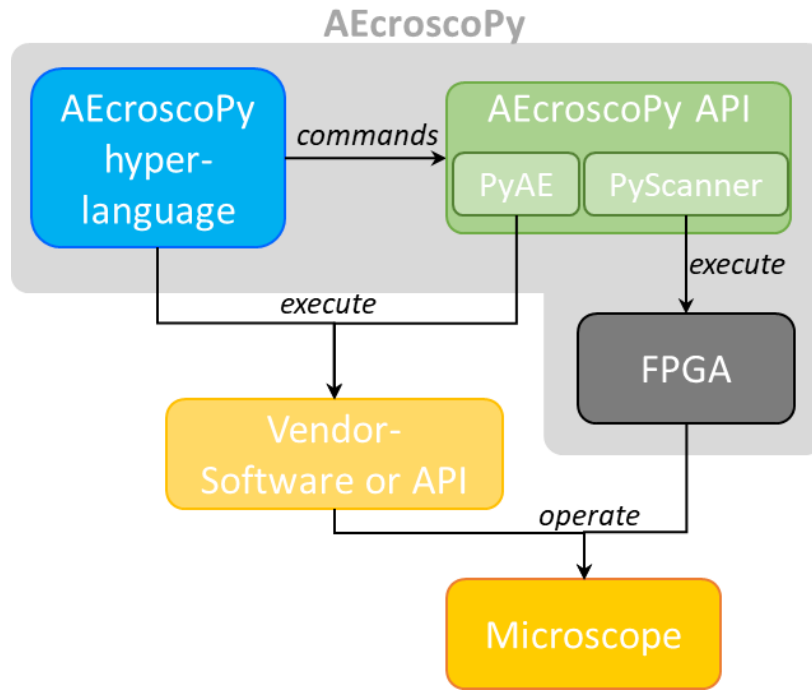

**Figure 1.** AEcroscoPy- Microscope system for automated and autonomous experiments.

## 2. Experiment and data management

During all experiments, when calling any AEcroscoPy function, a logger is activated to log the command sent, and this log can be retrieved at the conclusion of the experiment to verify the tasks that were carried out, as shown in Figure 2a. In this way, the AEcroscoPy software package ensures reproducibility, via standardized data models (discussed below), and traceability, via data logging.

Standardization of data capture and processing occurs through the use of a standardized data model in the *sidpy* package, termed the *sidpy.dataset* object[21]. Briefly, this is an extension to a dask array object which contains information on the data type, the dimensions, and all associated meta-data. Moreover, the *dask* framework enables parallel processing. AEcroscoPy functions for acquisition return *sidpy.dataset* objects, and AEcroscoPy also allows for *sidpy.dataset* objects to be processed either locally or on more powerful servers, including on the cloud, as shown in Figure 2b-c. This is implemented by virtue of running a process server that awaits *sidpy.dataset* objects, reads them once they are sent, and performs the processing described by keywords in the metadata, before returning the *sidpy.dataset* object back to the instrument with the processed data and the metadata automatically saved.

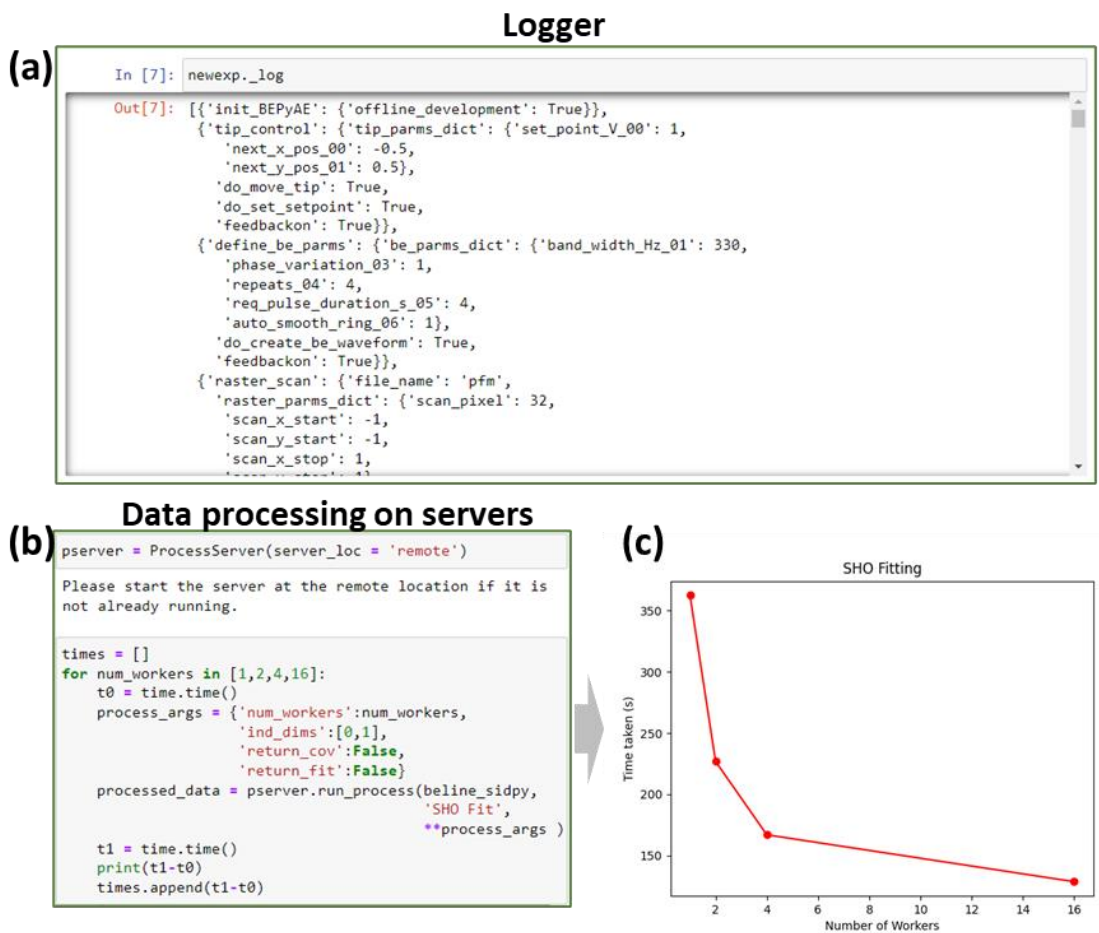

**Figure 2.** Experiment and data management in AEcroscoPy. (a) Logging capability automatically records all command sent to the microscope, which allows to track the experimental process. (b) users can also send the data to servers, running locally or remotely, for data processing. (c) using multiple cores on the server can accelerate the processing, with an example shown for fitting of spectra with a simple harmonic oscillator (SHO) model, where the data is captured on the instrument but processed elsewhere.

## 3. Showcase application in SPM

We show several illustrative examples of AEcroscoPy's application in SPM in Figures 2-4, highlighting AEcroscoPy's capacity of assisting researchers in high-throughput exploration of material manipulation, fine-tuning characterization parameters, and conducting spectroscopy measurements, respectively, in the context of Piezoresponse Force Microscopy (PFM), which is a technique for nanoscale characterization and manipulation of ferroelectric and piezoelectric materials. However, we note that the AEcroscoPy's application is not limited to these scenarios.

In PFM measurements, a conductive tip is brought in contact with the sample surface to apply an AC bias, inducing deformation in ferroelectric and piezoelectric materials through the inverse piezoelectric effect. It is noteworthy that various experimental parameters can influence the quality of PFM images and affect the materials under study. For instance, the AC bias determines the strength of the electric field applied to the sample and hence influences the signal to noise ratio, as well as potential to move the domain walls, while the set point governs the interaction force between the tip and the sample, which may lead to mechanical-induced polarization modulation (e.g., clamping) amongst other effects. Note that typically, larger setpoints are used to better nullify electrostatic contributions to the PFM signal.[18] Adjustment of these parameters not only allows us to acquire high-quality PFM images but also enhances our understanding of the intricate tip-material interaction during measurements. Traditionally, parameter optimization often requires manual adjustments by human operators to achieve the optimal measurement conditions or understand the tip-material interaction. However, AEcroscoPy enables us to systematically tune these parameters and analyze the results in a high-throughput manner automatically, reducing the need for extensive human labor.

Figure 3a presents PFM amplitude images of a $PbTiO_3$ thin film obtained with varying AC amplitude and set points. The amplitude image indicates the existence of ferroelastic and ferroelectric domains. The ferroelastic domains are seen as alternative dark and yellow stripes that are in-plane a-domains and out-of-plane c domains, respectively. The dark vertical line in the middle of the images is a ferroelectric 180º domain wall. The workflow for acquiring these results is documented in a Python Notebook, publicly available at Ref[22]. In brief, the workflow comprises two steps: parameter setting and image acquisition. The parameter-setting step picks a pair of AC voltage and set point values from a pre-established list, passing them to BEPyAE for execution. Subsequently, in the image acquisition step, a Band-excitation PFM image is captured with these specified parameters. The workflow iteratively picks new AC voltage/set point pairs and subsequently acquires BE-PFM data. In Figure 3a, the AC voltage and set point values for each PFM image is labeled, where the AC voltage increases from left to right and the set point increases from bottom to top. Notably, the PFM measurement region contains a vertical 180° domain wall. An apparent observation in Figure 3a is the variation in the thickness of this domain wall when varying AC voltage and set point. Therefore, we calculated the wall thickness under various AC voltages and setpoints and present it as a heatmap in Figure 3b. It is seen that the thickest wall appears at an AC voltage of 4.0 V and a set point of 2.5 V, while the thinnest wall appears at an AC voltage of 0.5 V and a set point of 0.5 V. Due to poor image contrast under certain conditions, it is challenging to use a universal method to calculate wall thickness, so the region in the heatmap corresponding to these poor images (top-left corner) is left blank. We further analyzed the overall detected piezoresponse amplitude changes, as shown in Figure 3c, revealing an increase with rising AC voltage. There is also an intriguing observation of an abrupt drop in piezoresponse when the set point increases from 2.0 V to 2.5 V, which necessitates further investigation but may be due to clamping and/or changes in the contact mechanics. In addition, since the AC electric field and applied force (set point) can induce polarization changes, we see a shift of domain wall at the end of the experiment compared to its original position, shown in Figure 3d. The analysis of the domain wall location (Figure 3e) reveals a more pronounced shift of domain wall when AC voltage changes, perhaps indicative of irreversible domain wall motion which has been studied previously.[23, 24]

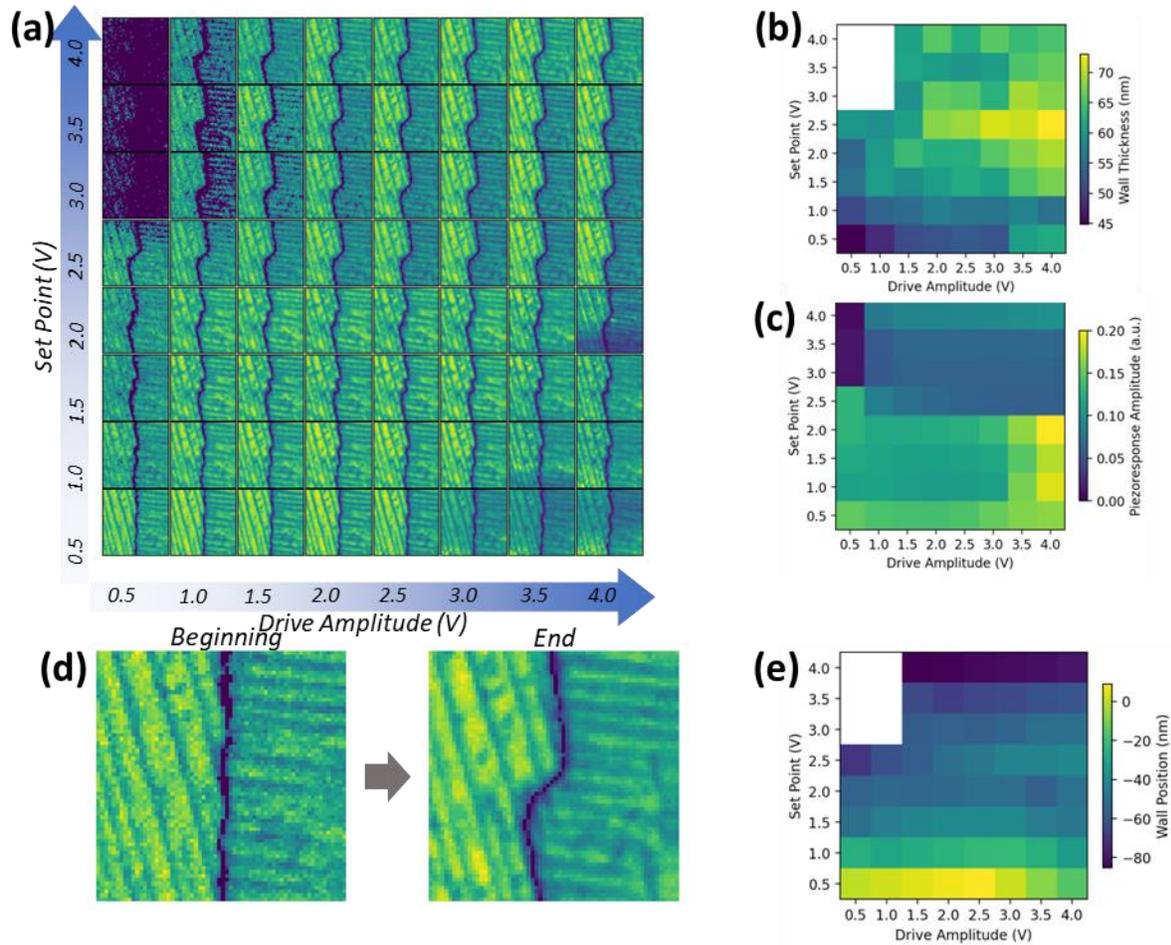

**Figure 3.** High-throughput experiment for investigating how PFM image parameters (i.e. drive amplitude and set point) affect ferroelectric domain wall thickness. (a) In this experiment, various drive amplitude and set point are used to image domain walls in a PTO thin film. The whole experiment is performed automatically with AEcroscoPy. (b) Post-experiment analysis show domain wall thickness as a function of drive amplitude and set point. (c) Overall piezoresponse of the image as a function of both setpoint and $V_{AC}$. (d) PFM measurements before and after indicating that the voltage and setpoints used modified the domain wall structure. (e) Wall position as a function of setpoint and $V_{AC}$.

In PFM, we can also switch the local polarization of a ferroelectric material $PbTiO_3$ thin film by applying a bias pulse to create domains with opposite polarization orientations, the created domains can be captured through a subsequent PFM image measurement. Often the bias pulse comprises two parameters: pulse magnitude and pulse duration. Investigating the relationship between the bias pulse and created domains allows to get insights into precise control of polarization switching and domain structures, essential to the application of ferroelectrics in data storage, sensors, actuators, etc. Figure 4 show results of a high-throughput exploration of bias pulse vs. domains in a PTO sample using AEcroscoPy. The workflow of this experiment comprises applying pulse to create domain and subsequent PFM image to capture created domains. Such a high-throughput approach allows us to study a large array of experiment conditions, offering systematic insights into the pulse-domain relationship. Shown in Figure 4e, we can observe that many domains grow alongside the ferroelastic wall. This can be interpreted as the pinning of 180º walls by ferroelastic walls, hence higher energy is required to promote the 180º walls to move across the ferroelastic walls. In addition, high-throughput experiment also allows to discover some phenomena that are rarely seen in traditional experiment (or they were ignored in traditional experiment because of rarity/lack of statistics). As shown in Figure 4f, we observe the double domains or triple domains when the applied pulse is centered on or near a ferroelastic domain wall.

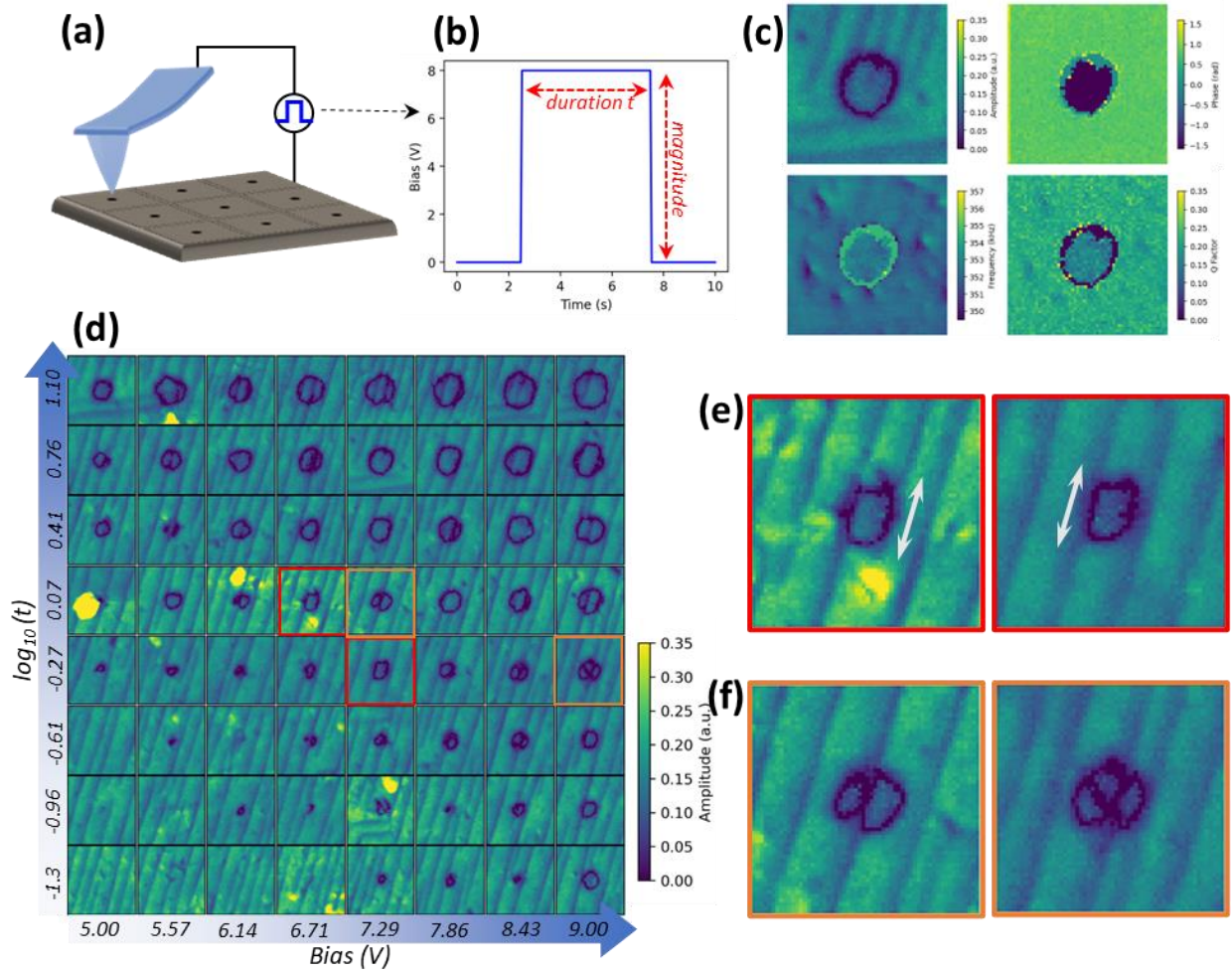

**Figure 4.** High-throughput experiment for ferroelectric domain writing, where various pulse conditions (i.e., pulse magnitude and duration) are used to write domain in a PbTiO$_3$ thin film, and a Band-Excitation PFM image measurement are performed to image the domain structure immediately after each pulse. The whole experiment is performed automatically with AEcroscoPy. The measurement setup is shown in (a) with the pulse parameters in (b). Example images from a single pulse are shown in (c), where the amplitude, phase, quality factor and resonant frequency are all plotted. (d) Full results (amplitude maps) and selected domains highlighted are shown in the colored insets in (e) and (f), indicating single, double and triple domain nucleation in the vicinity of ferroelastic walls.

In addition to standard PFM imaging modes that provide spatially resolved piezoresponse maps, the PFM spectroscopy mode enables investigation of the local electromechanical behavior as a function of an applied DC voltage. In PFM spectroscopy mode, we apply a voltage sweep and record the resulting piezoresponse of samples as a function of bias voltage at fixed locations on the sample. The PFM spectroscopy (i.e., hysteresis loop for ferroelectrics) offers information about ferroelectric characteristics such as remnant polarization, coercive field, and nucleation voltage at the nanoscale. PFM spectroscopy offers a deeper understanding of local electromechanical behavior, including imprint, fatigue, and domain nucleation/propagation. Traditionally, PFM spectroscopy measurements are performed either on a dense grid of points, or at specific locations of interest. The former case sacrifices some spatial resolution, takes a significant amount of time, and generated many measured points which are redundant. The latter case relies on experienced operators for many manual operations.

We often first scan the sample in standard imaging mode to identify specific locations of interest and mark these points. Then, move the PFM tip to the first marked point, apply voltage sweep and record the piezoresponse. Once the measurement at first point is completed, we move the tip to the next point and repeat the procedure. Further, we continue this process until all marked points are measured. This is time-consuming when all these operations need to be done manually. However, with AEcroscoPy, we can perform this PFM spectroscopy measurements at specific locations of interest automatically, as shown in Figure 5. As the first step, AEcroscoPy acquires a PFM image showing ferroelectric domains including in-plane a-domain (dark stripes in amplitude), out-of-plane c-domains (bright regions in amplitude), and domain walls. Here the c-domains are defined as the locations of interest, so a threshold filter is used on the image to extract the c-domains. As shown in Figure 5b, the white part is the extracted c-domains, then a certain number of points can be selected from the c-domains, either randomly or following specific rules. In Figure 4b, the blue spots indicate the locations selected from c-domain randomly, a few spots in a-domain (red spots) are also selected as a comparison. Then, a workflow written in AEcroscoPy can drive the tip to each location and trigger the spectroscopy measurements automatically. Shown as spectroscopy examples in Figure 5b are the averaged hysteresis loops from c-domains and a-domains, respectively, it indicates that the hysteresis loops from c-domains show larger remnant polarization than that of a-domains, which is consistent with known physics. Note that all these are done automatically without human-intervention using a workflow constructed with AEcroscoPy. Furthermore, we can also use AEcroscoPy to perform traditional grid spectroscopy measurement, an example 10x10 grid-spectroscopy measurement is shown in Figure 5c.

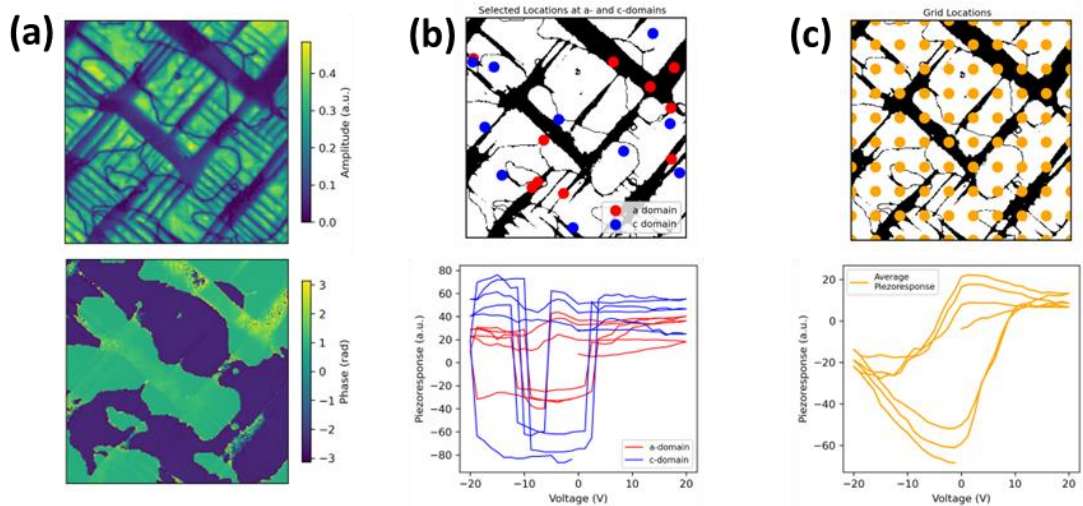

**Figure 5.** AEcroscopy enables band excitation piezoforce spectroscopy (BEPS) measurements at any specific locations (e.g., at a-domain or c-domain) or at grid locations, (a), PFM amplitude and phase images, (b) BEPS measurements at c-domain (blue spots) and a-domain (red spots), (c), BEPS at grid locations.

Additionally, we can use the AEcroscoPy platform to design and generate any specific scan trajectories to go beyond the traditional raster scan methods.[25, 26] The use of non-rectangular scans is advantageous for a variety of purposes: in functional SPM imaging can be used to increase time resolution,[27] or perform complex nanolithography processes; regarding hardware it can help reduce the high non-uniform accelerations that the XY piezo scanners suffer at the edges of the traditional raster scan trajectories, or reduce tip/sample damage; in the autonomous microscopy field it helps to obtain full control and versatility of the scanning parameters for experiment automatization which can be used to modify scan path on-the-fly or readjust for drift or region of interest tracking,[28] among others. In Figure 6, we show different scan trajectories obtained with PyScanner (see the corresponding output waveforms generated in python and input in the microscope controllers through the Pyscanner and FPGA in Supplementary information Figure S2). Examples in Figure 6a-d show traditional raster scan, spiral scan, Lissajous[29] scan and flower-like scan trajectories. Users can utilize these customized scan trajectories without programming, or they can program any customized scan path by uploading a NumPy array of their choice to control the X and Y scanners (or scan coils in the electron microscope).

When the density of the scan path is low (sparse scanning), specialized algorithms such as compressed sensing, gaussian process optimization or convolutional neural networks,[30] can be used to reconstruct the entire image from such sparse measurements in-painting the unscanned regions. For clarity purposes, we refer to the term *sparse* in its mathematical definition; that is an image (matrix) in which most measurement locations are zero (i.e., unscanned regions). In Figure 6e-g we show an example of Kelvin Probe Force Microscopy (KPFM) raster scan data, spiral scan raw data, and spiral scan reconstructed images that one can obtain using such scanning schemes. Note that the scan time required for a sparse scan is significantly lower than traditional raster scan. In addition to KPFM, such customized scan trajectories are applicable to any SPM characterizations, e.g., PFM, conductive AFM, etc. Below, we will also show the application of these customized scan trajectories in other microscopy techniques.

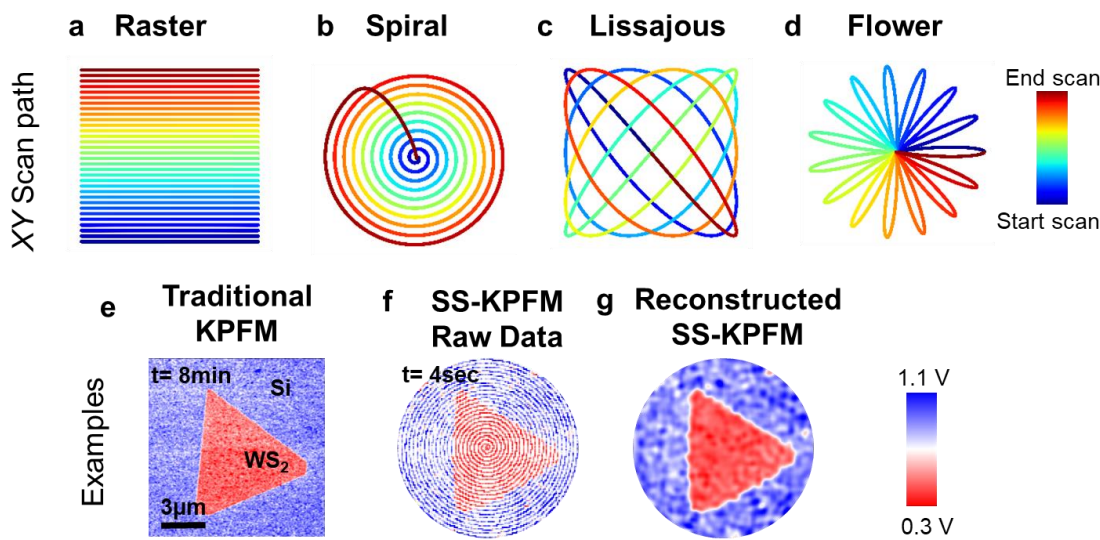

**Figure 6**. Customized scan trajectories: a) raster, b) spiral, c) Lissajous and d) flower. Note that users can design various scan trajectories in addition to the ones shown here. e-f) Example of KPFM data of $WS_2$ flake on Si acquired with raster and spiral scan trajectories (e) traditional raster scan strategy (f) sparse spiral scanning over the same sample region, (g) reconstructed map from sparse scan. Reproduced with permissions from [*R*ef, [27]].

**4. Application in other microscopy systems including NanoSurf, STEM.**

AEcroscoPy not only integrates a wide range of experimental setups, but its adaptability also extends to various instruments including atomic force microscopy (AFM), electron microscopy, and scanning tunneling microscopy. Above examples showcase AEcroscoPy application in the AsylumResearch Cypher AFM, but here we further highlight its applications on other microscopy platforms, such as NanoSurf AFM and electron microscopy.

Results in Figure 7 present PFM spectra on a $PbTiO_3$ thin film sample from a NanoSurf SPM, acquiring three spectra at 100 µm intervals. Here the sample movement after acquiring three spectra at each interval is managed by NanoSurf software built-in function and a motorized stage, AEcroscoPy orchestrates PFM spectroscopy measurements, mirroring the protocol as demonstrated in prior experiments. Figure 7 shows 30 PFM spectra over 10 different intervals, with a 50 nm spatial distinction between the 3 spectra at each interval. These results indicate the variation in hysteresis loops throughout the sample. AEcroscoPy allows for the flexibility on intervals, number of spectra at each interval, total number of the intervals, etc., tailoring to user specifications and interests. The demonstrated experiment here is especially useful for samples with pronounced spatial variations over large scale, such as in combinatorial libraries.[31] This broad scale experiment facilitates a systemic understanding of the properties relative to spatial parameters.

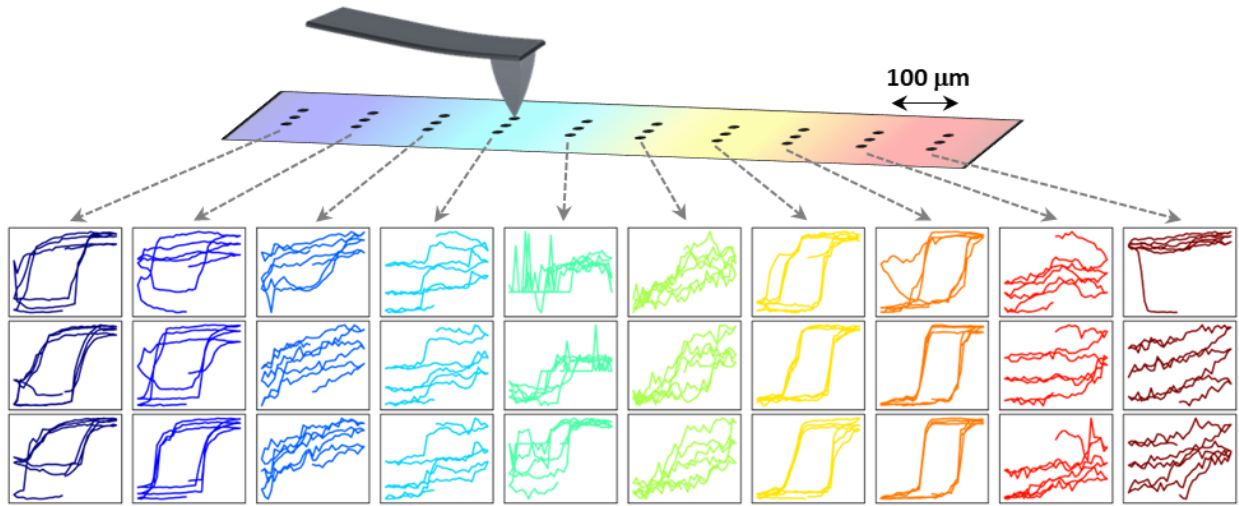

**Figure 7.** PFM spectroscopy measurement at NanoSurf AFM, where 3 PFM spectra apart 50 nm are obtained at each 100 μm interval.

Additionally, we can use the same software to capture spiral or other custom scan trajectories on the scanning transmission electron microscope. Specifically, the FPGA device is connected to a Nion UltraSTEM and used to directly control the scan coils. Shown in Figure 8 is an example of data collected in this fashion, for a single layer of suspended graphene, compared with the equivalent scanned in raster mode. It is clear, by observing the 2D FFT images, that the raster scan introduces a linear distortion that is absent in the spiral scan-acquired images. As such, it is a useful technique to not only control the overall beam dose (e.g., uniform dose, dose concentrated in outer radii, dose concentrated at center, and sparse scanned), but to also avoid artefacts arising from highly nonuniform accelerations on the X and Y controllers.

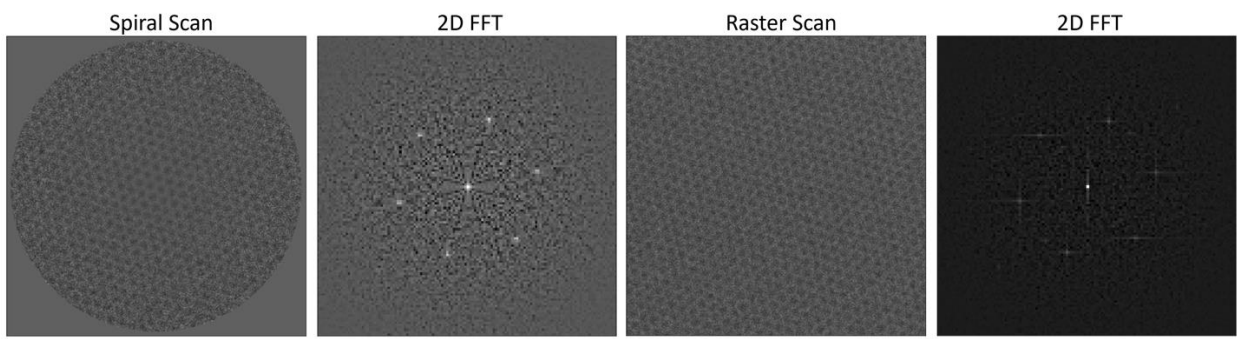

**Figure 8.** Spiral scan acquired image with 2D fast Fourier Transform (FFT) (left) and raster scan acquired image with 2D FFT (right) of a graphene sample in the scanning transmission electron microscope operated at 60 kV. The field of view of both images is 6 nm. The reflections in the 2D FFT of the spiral scan acquired image show clear, point-like features whereas those acquired from the raster scan show linear artefacts, which arise primarily from scan artifacts along the slow scan axis.

## 5. Integrating other tools within microscopy experiments.

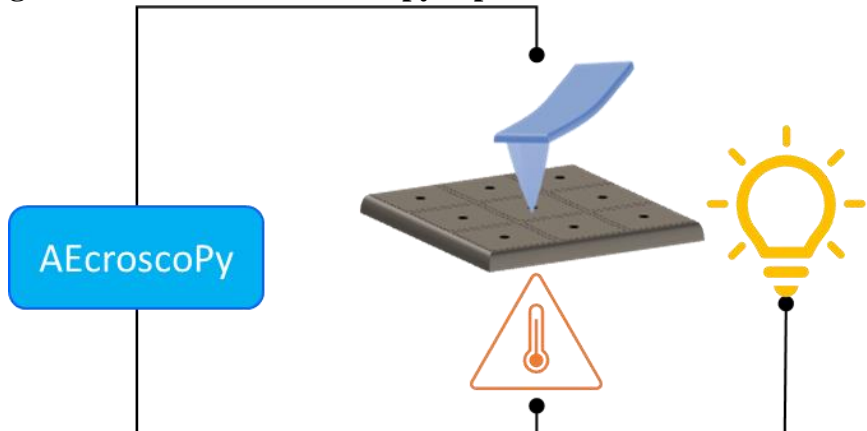

**Figure 9.** AEcroscoPy enables orchestration of other tools alongside with microscope, such as light and/or temperature source.

AEcroscoPy also allows the simultaneous operation of multiple tools alongside microscopy, as illustrated in Figure 9. This enables the design and execution of intricate experimental setups that were largely infeasible previously. For example, factors like temperature and light play significant roles in physical properties of samples, such as phase, mobility, strain, and charge density. Varying these factors in real-time enables capturing these dynamic processes. Typically, probing properties relative to these factors predominantly depended on instrument intrinsic sources, such as built-in light sources and temperature stages in microscopes. Hence, experimental designs are largely limited by the incorporated resources, which renders certain experiments infeasible, e.g., we are not able to investigate samples that are photoactive under UV light with a microscope built with a visible-range light source. However, AEcroscoPy allows orchestrating external tools that can be controlled via four channels in the FPGA, transcending these limitations. This opens the door for a spectrum of experiment that were not possible previously. Looking ahead, the orchestration of multiple scientific instruments promises further revolution of experimental research, towards the 'smart autonomous research facilities' concept.[8]

## 6. Achieve ML-driven microscopes by AEcroscopy

AEcroscoPy also facilitates the seamless integration of machine learning (ML) algorithms into microscopy experiments for accelerated discoveries and augmented analytical precision. Figure 9 showcases several examples of integrating ML within microscopy workflows using AEcroscoPy.

The first example describes a scenario where a human expert defines a feature of interest and ML identifies this feature within the image data, then an AEcroscoPy constructed workflow drives the microscope to delve deeper into this feature. Similar experiments have been reported by us.[32, 33] In this scenario, image data capture is achieved using *raster_scan()* function in AEcroscoPy. Once ML algorithm identifies the feature, *do_BEPS_specific()* drives the microscope to the location of the feature for PFM spectroscopy measurement to investigate detailed response of this feature.

The second instance is the search for a user-defined target property discerned from spectroscopy data. The experiment begins with PFM spectroscopy measurement at random locations using *do_BEPS_random()* function. Then, users evaluate these spectra and define the target spectrum. Next, a Gaussian process Bayesian Optimization algorithm predicts the next location with the highest likelihood of the target spectrum, subsequently *do_BEPS_specific()* function performs the PFM spectral measurement at this location. We refer our earlier publication for readers interested in this experiment.[34]

The third scenario is to explore structure-property relationships. In this scenario, the initial step is to acquire image data with the *raster_scan()* function, which provides nanoscale structure within the sample. Then, a few spectral can be obtained either at random locations using *do_BEPS_random()* or at predetermined locations using *do_BEPS_specific()*. Specific ML algorithms then delve into analyzing the relationship between structure and spectra. Further, ML predicts the optimal location for further spectral measurement to refine this established relationship. The *do_BEPS_specific()* function will undertake the next spectral measurement. Prior works on the exploration of structure-property relationships can be found elsewhere, our focus here is to touch upon the capability of AEcroscoPy for harmonizing ML with microscopy experiments instead of in-depth exploration of ML algorithms or specific experiments, while we refer previous works if readers are interested in.[35-40]

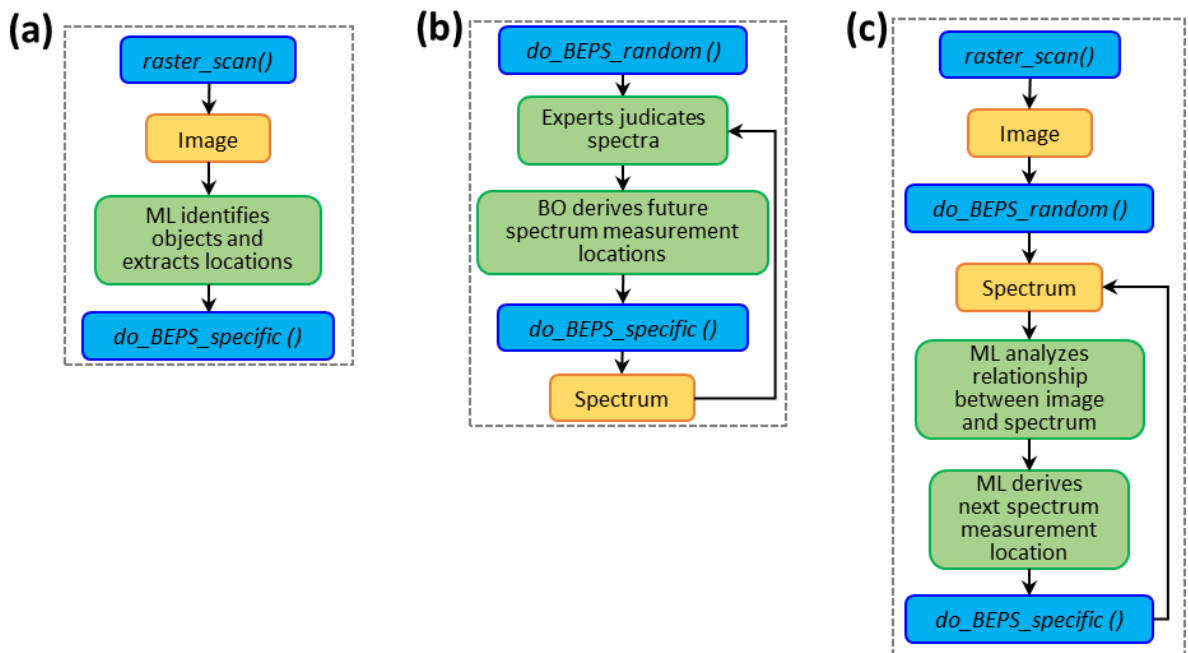

**Figure 9.** Examples of seamless integration of ML algorithms with microscope workflows using AEcroscoPy.

**Conclusion**

In summary, AEcroscoPy as a platform synergizes hardware and software to overcome limitations of human-centric operations not only automates microscope experimentations but also broadens the scope, utility, efficiency, effectiveness, and reproducibility of microscope experiments. We showcased representative examples of utilizing AEcorscoPy in various microscopes including atomic force microscopy, scanning tunneling microscopy, and scanning transmission electron microscopy. However, the application of AEcroscoPy extends to other microscope systems as well, such as scanning electron microscopy cathodoluminescence. In addition to experimental automations, AEcorscoPy also excels in performing rapid corrections or feedback at speeds unattainable by human operators, and excels in monitoring microscope conditions over multiple experiments. These capabilities make AEcroscoPy be readily integrated with machine learning and simulations for autonomous experimentation. At the end, we recommend readers visit AEcroscoPy website at https://yongtaoliu.github.io/aecroscopy.pyae/welcome_intro.html for the latest information and updates when using AEcroscoPy.

### Acknowledgements
This research was supported by the Center for Nanophase Materials Sciences (CNMS), which is a US Department of Energy, Office of Science User Facility at Oak Ridge National Laboratory.

### Conflict of Interest
The authors declare no conflict of interest.

### Authors Contribution
Y.L. and R.K.V developed AEcroscoPy python package. S.J. developed PyAE and PyScanner software. S.M.V. assisted in coding parts of AEcroscoPy. Y.L. constructed the AEcroscoPy tutorial webpage and collected AFM results. M.C. collected customized scan trajectories AFM results and K.R. collected customized scan trajectories STEM results. All authors contributed to discussions and the final manuscript.

### Data Availability Statement
The method that supports the findings of this study are available at https://yongtaoliu.github.io/aecroscopy.pyae/welcome_intro.html.